\documentclass[12pt,amsmath,amssymb]{revtex4}
\usepackage{graphicx}% Include figure files
\usepackage{dcolumn}% Align table columns on decimal point
\usepackage{bm}% bold math
\usepackage{multirow}
\usepackage{booktabs,siunitx}

\begin{document}

\preprint{APS/123-QED}%

\title{Exchange bias in Sm$ _{2} $NiMnO$  _{6}$/BaTiO$  _{3}$ ferromagnetic-diamagnetic heterostructure thin films}% Force line breaks with \\

\author{S. Majumder}
\author{S. Chowdhury}
\author{B. K. De}
\author{V. Dwij}
\author{V. Sathe}
%\author{$^{c}$}
%\author{$^{c}$}
\author{D. M. Phase}
\author{R. J. Choudhary}\email{ram@csr.res.in}

\affiliation{UGC DAE Consortium for Scientific Research, Indore 452001, India}
%\affiliation{$^{b}$Beamline Development and Application Section, Bhabha Atomic Research Centre, Mumbai 400085, India}
%\affiliation{$^{c}$Elettra Sicrotrone Trieste S.C.p.A., SS 14-km 163.5, 34149 Basovizza, Italy}


\begin{abstract}
Exchange bias (EB) shifts are commonly reported for the ferromagnetic (FM)/antiferromagnetic (AFM) bilayer systems. While stoichiometric ordered Sm$_{2}$NiMnO$_{6}$ (SNMO) and BaTiO$_{3}$ (BTO) are known to possesses FM and diamagnetic orderings respectively, here we have demonstrated the cooling field dependent EB and training effects in epitaxial SNMO/BTO/SNMO (SBS) heterostructure thin films. The polarized Raman spectroscopy and magnetometric studies reveal the presence of anti-site cation disorders in background of ordered lattice in SNMO layers, which introduces Ni-O-Ni or Mn-O-Mn local AFM interactions in long range Ni-O-Mn FM ordered host matrix. We have also presented growth direction manipulation of the degree of cation disorders in the SNMO system. Polarization dependent X-ray absorption measurements, duly combined with configuration interaction simulations suggest charge transfer from Ni/Mn 3\textit{d} to Ti 3\textit{d} orbitals through O 2\textit{p} orbitals across the SNMO/BTO (SB) interfaces, which can induce magnetism in the BTO spacer layer. The observed exchange bias in SBS heterostructures is discussed considering the pinning of moments due to exchange coupling at SB (or BTO/SNMO) sandwich interface.
\end{abstract}

\maketitle

%\section{INTRODUCTION}
\noindent
The exchange bias (EB) phenomenon, described as the shift of isothermal magnetic hyteresis loop in the magnetic field scale, is attributed to unidirectional anisotropy effect \cite{CWang2018, JVKima2001}. Ideally an antiferromagnetic (AFM) system possess equivalent energies for the two oppositely aligned magnetic sublattices. Now, the EB effect can be understood considering the exchange coupling of any one AFM sublattice spins with adjacent ferromagnetic (FM) spin arrangements, when the FM/AFM interface system (FM T$_{C} > $ AFM T$_{N}$) is cooled in presence of magnetic field from temperature T$>$T$_{N}$. This will cause breaking of energetic equivalence of the AFM sublattices, which eventually results in requirement of an additional magnetic field for the reversal of magnetization direction of the FM spins overcoming the coupling interaction. The sign of this bias field depends on the sign of cooling field \cite{JNogues1996}. The magnitude of EB field, defined as H$ _{EB} $ = [($|$H$_{C+}| - |$H$_{C-}|$)/2], is governed by the moment pinning strength at the interface separating the FM and AFM phases. Materials possessing EB character have huge relevance in technological usages, for instance: spin valve, magnetic random access memory, magnetoresistive and recording devices etc \cite{SGider1998, BDieny1991, VKuncser2008, ALOrtega2015}. 

In addition to conventional FM/AFM composite systems, EB properties are also reported for hard-FM/soft-FM, FM/ferrimagnetic, FM or AFM /spin glass, FM/paramagnetic (PM) interfaces etc. \cite{EManiv2021, SGiri2011, JSort2004, CBinek2006, TMaity2013}. On the other hand, structural defects can induce or modify EB by introducing moment pinning centers in the system, leading to local alteration in magnetic exchange and anisotropy \cite{CWang2018, JVKima2001, SGBhat2017, RCSahoo2019, MDas2020}. In the present work, we plan to investigate the EB effect in Sm$_{2}$NiMnO$_{6}$/BaTiO$_{3}$/Sm$_{2}$NiMnO$_{6}$ (SBS) heterostructure. Assuming the ideal scenario (defect free crystal structure) Sm$_{2}$NiMnO$_{6}$ (SNMO) and BaTiO$_{3}$ (BTO) can be categorized as FM and diamagnetic systems due to Ni$ ^{2+/3+} $-O-Mn$ ^{4+/3+} $ near linear super exchange interaction and Ti$ ^{4+} $ 3\textit{d}$ ^{0} $ configuration, respectively \cite{SMajumder2022prbb, SMajumder2019}. However, in our previous studies we have observed that presence of defects in SNMO and BTO host matrix has huge bearings on their corresponding magnetic behaviors \cite{SMajumder2022prbt, SMajumder2019}. Here, we have demonstrated how defect induced alteration in electronic structure can drive unusual magnetic behavior like EB phenomenon in SBS heterostructure system. 

In order to explore the role of different growth directions and strain states, the SNMO $\sim$75$\pm$5 nm/BTO $\sim$150$\pm$5 nm/SNMO $\sim$75$\pm$5 nm heterostructure thin films were fabricated simultaneously on (001) oriented LaAlO$ _{3} $ (LAO), (001) oriented SrTiO$ _{3} $ (STO), and (111) oriented STO substrates and named as SBS$\_$L001, SBS$\_$S001, and SBS$\_$S111, respectively. Following the previously optimized recipe as reported in Refs. \cite{SMajumder2022prbt, SMajumder2019}, here we have employed 700 $^{\circ}$C of deposition temperature (DT) and 500 mTorr of oxygen partial pressure (OPP) to grow the SBS systems. The crystal structure and phase information of the grown thin films were investigated utilizing Cu K$ \alpha $ ($ \lambda$ = 1.54 $ \AA $) radiation X-ray diffractometers (Bruker D2 Phaser, high resolution Bruker D8 Discover), and linearly polarized blue laser ($ \sim $ 473 nm) source equipped micro-Raman spectrometer (Horiba Jobin-Yvon). The magnetic properties were studied using MPMS 7-Tesla SQUID-VSM (Quantum Design Inc., USA) instrument, which has average sensitivity of the order of $ \sim 10^{-8}$ emu in determining magnetic moments. The electronic interactions in the samples were explored by total electron yield (TEY) X-ray absorption near edge spectroscopy (XANES) experiments with soft X-ray beam linearly polarized along horizontal direction to the synchrotron storage ring (SXAS beamline, Indus 2, BL 1, RRCAT, Indore, India). The incident beam energy calibration in XANES spectra was performed by aligning the measured edge position of the reference standard to known absorption energy value and the estimated maximum energy resolution was found to be $ \sim $ 0.25 eV within the studied energy range. For the analysis of transition metal XANES spectra we have used CTM4XAS \cite{HIkeno2009} simulations.

\begin{figure*}[t]
\centering
\includegraphics[angle=0,width=0.95\textwidth]{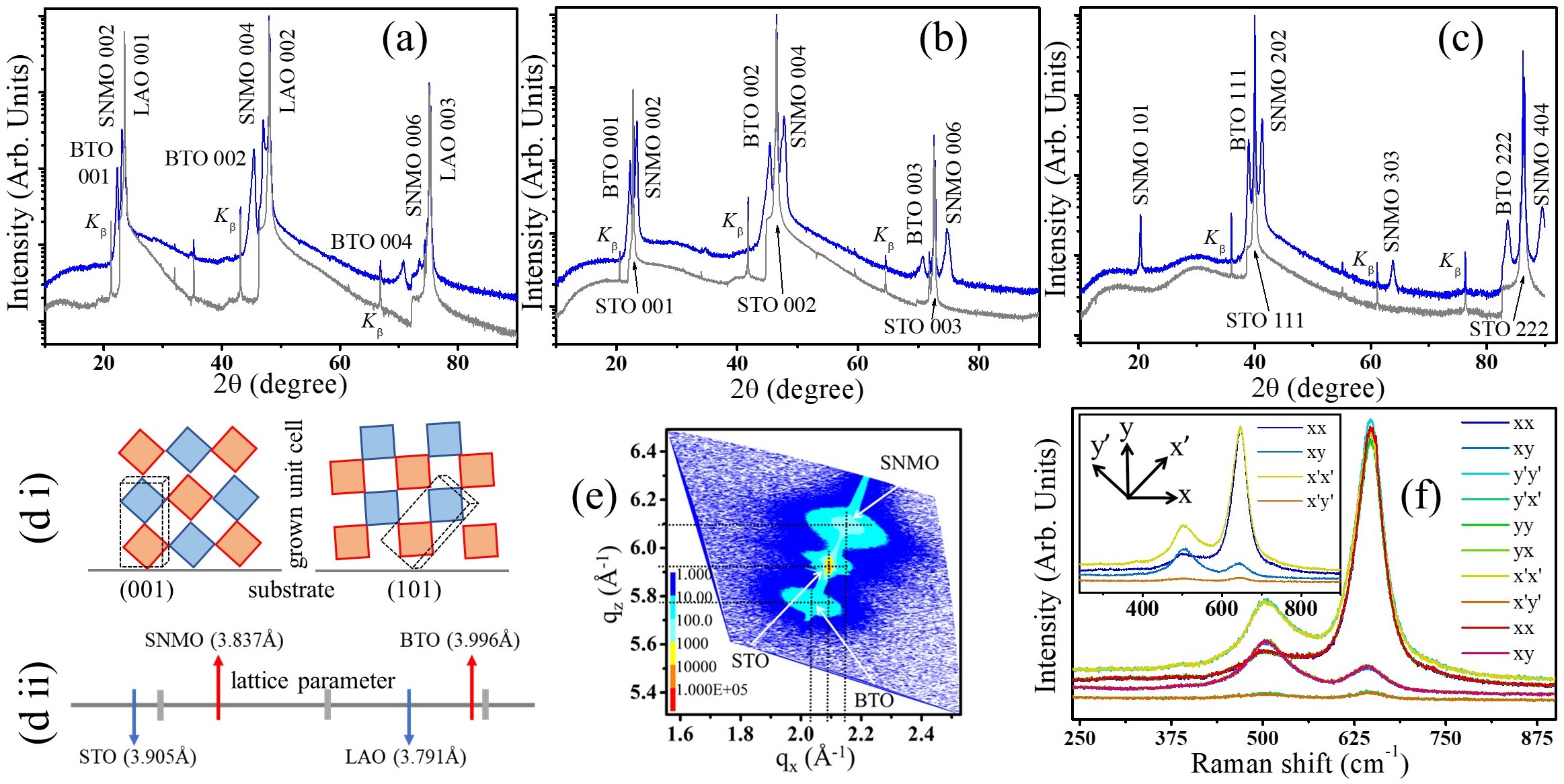}
\caption{X-ray diffraction $2\theta $ scans of SNMO/BTO/SNMO heterostructure thin film samples grown on top of (a): LAO (001), (b): STO (001), and (c): STO (111) substrates. To have better visualizations XRD patterns are vertically translated here. Peaks at the left side to that of the substrate peaks are assigned to K$ _{\beta} $ reflections. Schematic representations for (d i): different growth directions dependent octahedral arrangements for different oriented substrates, (d ii): different lattice mismatch strains for different substrates. (e): Representative asymmetric reciprocal space map around (121) plane of the SBS$ \_ $S111 thin film. The most intense coordinates correspond to the SNMO, BTO layers and STO substrate are highlighted by arrows in (q$ _{x} $, q$ _{z} $) domain. (f): Polarized Raman spectra recorded in different angular configurations as shown the measurement geometry in the Inset.}\label{xrdrsmraman}
\end{figure*}

The X-ray diffraction (XRD) 2$ \theta $ scans of the grown samples SBS$ \_ $L001, SBS$ \_ $S001, and SBS$ \_ $S111 are depicted in Figs. \ref{xrdrsmraman}(a-c), respectively. All the heterostructures exhibit single phase and oriented XRD peaks corresponding to SNMO monoclinic (\textit{P2$  _{1}$/n}) or orthorhombic (\textit{Pbnm}) and BTO tetragonal (\textit{P}4\textit{mm}) crystal symmetries. Depending on the substrate orientation, the growth direction of the heterostructure changes, which has significant impact on defect formation in the system (discussed later). We have schematically illustrated this in Fig. \ref{xrdrsmraman}(d i). The lattice parameters of bulk SNMO, BTO, LAO, and STO crystal systems, approximated to average pseudo-cubic values, are presented in Fig. \ref{xrdrsmraman}(d ii). These lattice mismatch between thin film and substrate materials may cause misfit strain on the grown over layer crystals. A representative reciprocal space mapping across the asymmetric (121) plane for the SBS$ \_ $S111 sample is shown in Fig. \ref{xrdrsmraman}(e). RSM scans reveal that despite of epitaxial growth, SBS heterostructures are relaxed from the substrate induced strains, and consequently the estimated unit cell parameters in SNMO, BTO layers are found close to its bulk counterparts. The stain relaxation indicates incorporation of structural defects such as edge dislocation, in the grown heterostructures. It is worth to mention here that in BTO and SNMO systems the oxygen vacancy and anti-site cation occupancy defects respectively, have low formation energies and even can be manipulated by controlling the fabrication process \cite{SMajumder2022prbt, SMajumder2019}. 

To probe the microstructure of the SNMO layers in SBS heterostructure, we have utilized angle dependent polarized Raman spectroscopy measurements, as shown in Fig. \ref{xrdrsmraman}(f) for SBS$\_$L001 sample. The spectral scans were recorded in different geometrical configurations, where the two letters (for example: xy) notation corresponds to the incident (x) and scattered laser lights polarization directions (y), as depicted by the schematic in Fig. \ref{xrdrsmraman}(f). At first glance, two distinct Raman peaks centered at $ \sim $503, 647 cm$ ^{-1} $ are observed, which are assigned as antistretching (bending) and stretching (breathing) vibrational motions of (Ni/Mn)O$ _{6} $ octahedra, respectively \cite{MNIliev2007}. As the Ni/Mn cation ordered and disordered phases belong to different space groups, the Raman modes follow different polarization selection criteria. For ordered structure (\textit{P2$  _{1}$/n}), the stretching spectral feature obeying \textit{A}$ _{g} $ symmetry is allowed in xx, x'x', yy and forbidden in xy scattering geometries. Whereas for disordered structure (\textit{Pbnm}), the stretching spectral feature obeying \textit{B}$ _{1g} $ symmetry is allowed in xy, x'x' and forbidden in xx, yy, x'y' configurations \cite{MNIliev2007}. In the observed spectra, the stretching Raman mode showing spectral intensity for both xx, xy polarizations, confirms coexistence of major ordered and minor disordered structures of SNMO in SBS system. Furthermore, the observance of distinguishable differences in xx, x'x' and xy, x'y' Raman spectra, again suggests the epitaxial nature of the grown sample. 

\begin{figure*}[t]
\centering
\includegraphics[angle=0,width=0.95\textwidth]{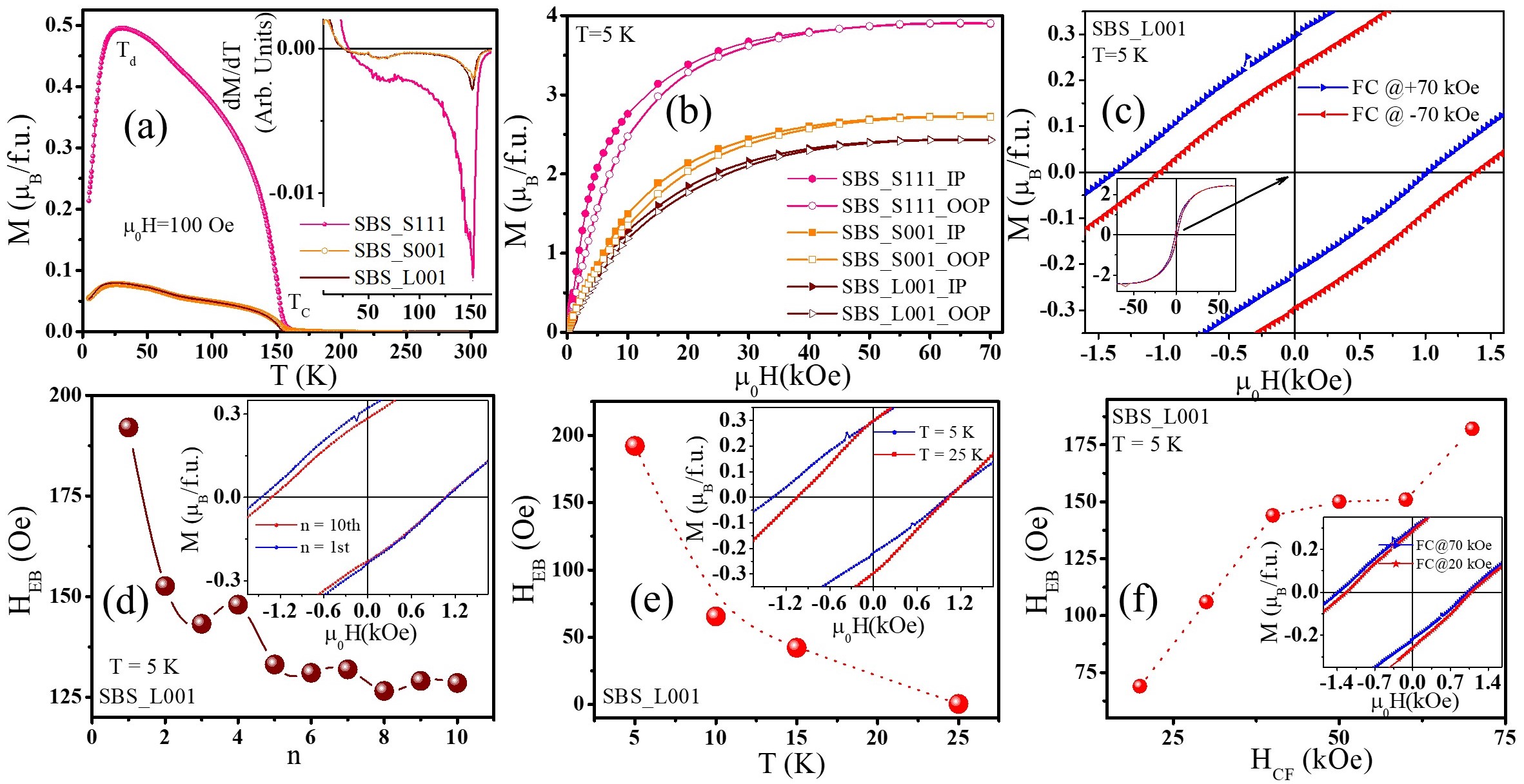}
\caption{Temperature dependent magnetization M(T) recorded in field cooled warming cycle with magnetic field $ \mu_{0} $H=100 Oe for SBS heterostructures. Inset shows first order temperature derivative of magnetization dM(T)/dT. (b): Zero field cooled isothermal (T=5 K) magnetic virgin curves measured as a function of applied magnetic field M(H) along in-plane and out-of-plane geometric configurations. Field cooled exchange bias at 5 K temperature for the SBS$\_$L001 sample. Insets show corresponding full scale M(H) data. For the SBS$\_$L001 sample, variation of exchange bias field (d): with repetitive training cycles, (e): for different measuring temperatures, and (f): with varying cooling fields. Insets of each plots show enlarged view of the low field regions of corresponding M(H) results.}\label{mtvceb}
\end{figure*}

In order to ascertain the magnetization behavior of the grown heterostructures, we have performed dc magnetometric measurements as a function of temperature M(T) and in isothermal condition by varying applied magnetic field M(H). The field cooled warming cycle of M(T) recorded in presence of 100 Oe external magnetic field as presented in Fig. \ref{mtvceb}(a), shows two prominent transitions: (i) the onset in moment at T$ _{C} $=151.5 K for SBS$\_$L001, 152.2 K for SBS$\_$S001, 152.4 K for SBS$\_$S111; and (ii) the downturn in moment at T$ _{d} $=24.9 K for SBS$\_$L001, 26.9 K for SBS$\_$S001, 30.6 K for SBS$\_$S111. Point to be noted here that the BTO spacer layer, assuming to be stoichiometric, should not contribute any magnetic signal other than a diamagnetic background due to its Ti 3\textit{d}$ ^{0} $ electronic configurations \cite{SMajumder2019}. The transition observed at T=T$ _{C} $ is assigned as the paramagnetic (PM) to long range FM ordering due to Ni-O-Mn superexchange interactions in cation ordered Ni-Mn network \cite{SMajumder2022prbb, SMajumder2022prbt}. The transition across T=T$ _{d} $ is related to the opposite arrangement of polarized Sm PM moments with respect to Ni-Mn FM sublattice \cite{SMajumder2022prbb, SMajumder2022prbt}. A broad dip like feature in M(T) within T$_{d}<$T $<$T$_{C}$, which is identified as the signatures of Ni/Mn cation disorders \cite{SMajumder2022prbt}, is prominently observed for SBS$\_$L001 and SBS$\_$S001 samples. Comparing the M(T) results for different SBS samples, the increment in transition temperatures and vanishing trend of dip feature indicate better cation ordering in SBS$\_$S111 than SBS$\_$L001 or SBS$\_$S001 samples. 

From the zero field cooled magnetic isotherms acquired at T=5 K, as depicted in Fig. \ref{mtvceb}(b), obtained saturation moment (at $ \mu_{0} $H=70 kOe) values are found to be M$ _{S} $=2.4 $\mu_{B}$ for SBS$\_$L001, 2.7 $\mu_{B}$ for SBS$\_$S001, and 3.9 $\mu_{B}$ for SBS$\_$S111. Magnetic virgin curves recorded with applied magnetic field along the in-plane (IP) and out-of-plane (OOP) directions, are used to estimate the the effective anisotropy as \begin{equation}\label{Eqanisotropykeff}
\kappa_{eff} = \int_{0}^{\mu_{0}H_{S}}M(H)_{in} dH - \int_{0}^{\mu_{0}H_{S}}M(H)_{out} dH
\end{equation}
where H$ _{S} $ is saturation field. The obtained effective anisotropy energies are found to be $\kappa_{eff}$=2.548 $ \times $ 10$ ^{5} $ erg/cm$ ^{3} $ for SBS$\_$L001, 3.061 $ \times $ 10$ ^{5} $ erg/cm$ ^{3} $ for SBS$\_$S001, and 5.299 $ \times $ 10$ ^{5} $ erg/cm$ ^{3} $ for SBS$\_$S111. Ni/Mn cation disorders introduce Ni-O-Ni, Mn-O-Mn short scale AFM interactions in the background of ordered FM host matrix and consequently, intercompeting magnetic exchanges coexists in SNMO for a wide temperature range \cite{SMajumder2022prbb, SMajumder2022prbt}. The decrease in saturation moments and anisotropy energy again confirms the progressive improvement in chemical ordering as: SBS$\_$L001 $ < $ SBS$\_$S001 $ < $ SBS$\_$S111. The SNMO layers grown on S111 substrate have a better cation ordering as compared to S001 substrate due to its preferential alternate arrangements of Ni and Mn octahedra when stacked along IP than OOP directions.

\begin{figure*}[t]
\centering
\includegraphics[angle=0,width=0.95\textwidth]{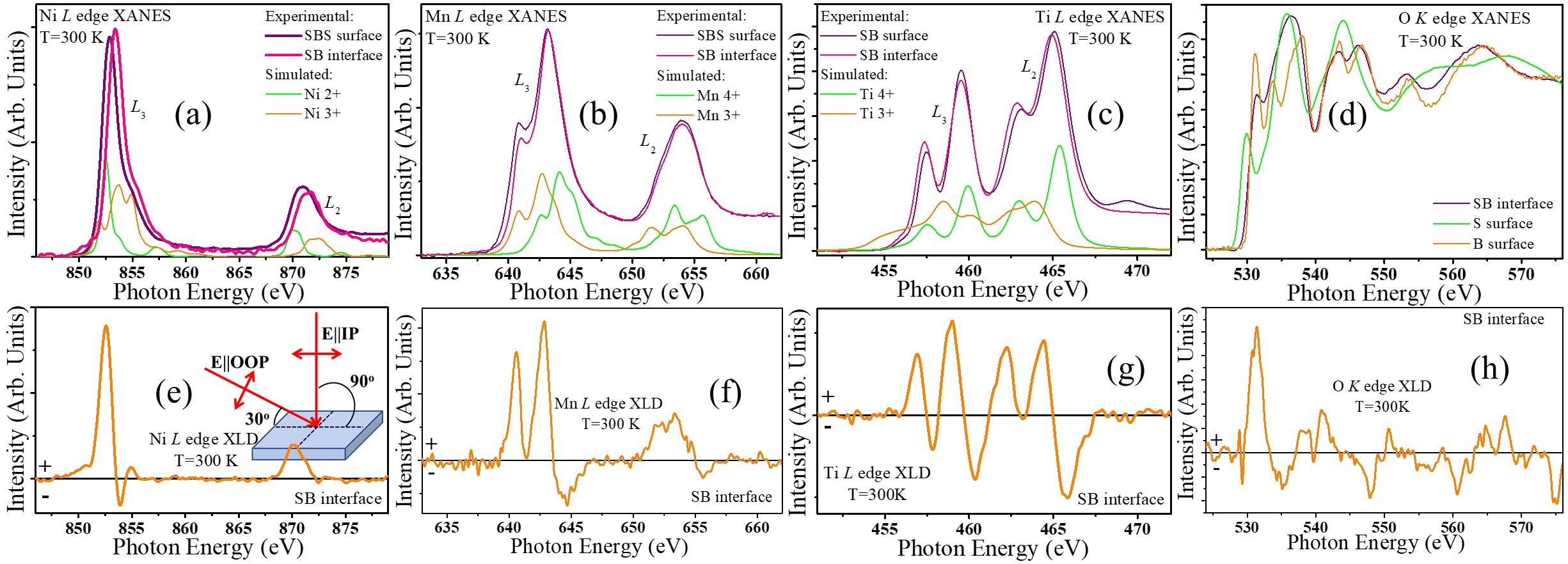}
\caption{Total electron yield mode detected experimental and configuration interaction charge transfer multiplet simulated isotropic X-ray absorption spectra for the SNMO/BTO hetero-interface on LAO (001) substrate measured across the (a): Ni, (b): Mn, and (c): Ti \textit{L} edges. (d): Isotropic O \textit{K} edge XANES for SB hetero-interface, surfaces of SNMO, BTO stoichiometric samples. Inset of (e) show schematic representation for the X-ray linear dichroism measurement geometry. XLD spectra for the SNMO/BTO hetero-interface on LAO (001) substrate measured in TEY mode across the (e): Ni, (f): Mn, (g): Ti \textit{L}, and (h): O \textit{K} edges.}\label{xanesxld}
\end{figure*}

The negative (or positive) field cooled (FC) M(H) hysteresis loops (under 70 kOe magnetic field at T=5 K temperature), as depicted by a representative plot in Fig. \ref{mtvceb}(c) for SBS$\_$L001 sample, reveal EB shift toward positive (negative) field and negative (positive) moment scales. The EB field, averaged over positive and negative FC cycles, are found to be $ \sim $182 Oe for SBS$\_$L001, 91 Oe for SBS$\_$S001, and 85 Oe SBS$\_$S111. To avoid possible artifact in EB related to the minor loop effects, in all these M(H) measurements the scanning field was kept above the saturation field $ \sim $55 kOe. For further characterization of EB, characteristic training behaviors are explored, as shown in Fig. \ref{mtvceb}(d) for sample SBS$\_$L001. The observed gradual decrements in EB field with repeated magnetic field cycling suggests change in anisotropy at the moment pinning boundaries with each successive cycle \cite{RCSahoo2019}. As the measuring temperature is increased, EB is found to reduce, as shown in Fig. \ref{mtvceb}(d) for SBS$\_$L001 sample. This is because of dilution of AFM ordering strength with rising system temperature \cite{JNogues1996}. Keeping the scanning field fixed at $ \pm $70 kOe, EB is traced for different cooling fields, as presented in Fig. \ref{mtvceb}(e) for SBS$\_$L001 sample. This increase in EB with increasing cooling field results from the consequences of external magnetic field during the cool down process on the pinning mechanism at the interface involed with FM \ AFM layers, \cite{JNogues1996}. It is important to mention here that we have not observed any EB shift in similarly disordered individual SNMO thin film (data not shown here). This is possibly because in SNMO system, the disordered bonds of short range AFM interactions are randomly distributed in the ordered host matrix of FM network, and therefore the FM/AFM interfacial coupling is not sufficient to pin the moments for EB. This indicates that BTO spacer layer contributes in magnetism of SBS heterostructure, and liable for the observed EB effect. Now in the present scenario the induced magnetism in BTO layer may have two possible ways: (i) charge transfer across the SNMO/BTO interface, and (ii) interfacial oxygen vacancy in BTO layer. 

To examine the electronic structure origin of observed magnetic behavior in SBS system we have measured X-ray isotropic absorption spectra, as depicted by the representative plots in Fig. \ref{xanesxld}(a-d) for the L001 heterostructure system. Comparing the experimentally observed and configuration interaction simulated Ni and Mn 2\textit{p} $ \rightarrow $ 3\textit{d} excitation (\textit{L}) edge isotropic XANES spectra (Fig. \ref{xanesxld}(a, b)), it is confirmed that in SBS heterostructure surface both Ni and Mn have mixed valency in between 2+, 3+ and 4+, 3+, respectively. As the soft X-ray absorption measurement probing depth is confined to $ \gtrsim $5 nm from the surface of the sample, the aforementioned XANES study on SBS sample provides the information from the SNMO surface layer only. Similarly, at the SNMO / BTO (SB) surface with $ \backsim $75 nm of BTO top layer on SNMO bottom layer, the Ti ions mostly hold 4+ oxidation state (Fig. \ref{xanesxld}(c)). Now, to elucidate the induced magnetism in BTO, we have investigated XANES at the interface of SNMO / BTO (SB) with $ \backsim $1 nm BTO top layer on SNMO bottom layer. As we move from the SBS surface (thick SNMO top layer) to SB interface (ultralow BTO top layer on SNMO), the Ni/Mn \textit{L} edge positions show small shift towards higher energy side, indicating slight increase in charge state of Ni/Mn cations. On the other hand, when we move from the SB surface (thick BTO top layer) to SB interface (ultralow BTO top layer on SNMO), Ti \textit{L} edge positions show small shift towards lower energy side, indicating slight decrease in charge state of Ti cations. These observations point out the possible charge transfer from Ni/Mn ions to Ti ion across the interface which can transform Ti$ ^{4+} $ into Ti$ ^{3+} $.

To have a further insight of this change transfer picture, we have measured X-ray linear dichroism (XLD) following the geometry as shown schematically in the Inset of Fig. \ref{xanesxld}(e). Absorption spectra recorded with incident photon beam polarized along the parallel (E$ \| \|$IP) and perpendicular (E$ \| \|$OOP) directions to the sample surface, preferentially probes the unoccupied states in the IP (\textit{d}$ _{x^{2}-y^{2}} $, \textit{d}$ _{xy} $) and OOP (\textit{d}$ _{3z^{2}-r^{2}} $, \textit{d}$ _{yz} $, \textit{d}$ _{xz} $) orbitals, respectively. The difference in these two spectra I$ _{IP} $ - I$ _{OOP} $, which estimates the orbital polarization, is known as XLD signal \cite{PCRogge2018}. At the SB interface Ni, Mn, and Ti all exhibit positive XLD signals, which indicate more unoccupied states in IP than OOP orbitals, or in other words more electron populations in OOP \textit{d} states. The O \textit{K} edge XANES spectral shape and intensity across $ \backsim $530 eV is a sensitive probe for the oxygen stoichiometry in the lattice system \cite{SMajumder2019}. However, the O \textit{K} isotropic XANES for the SB interface as presented in Fig. \ref{xanesxld}(d), is contributed from two different lattice environments of SNMO and BTO structures, and is inappropriate to examine the possible presence of interfacial oxygen vacancies. The O \textit{K} XLD shows preferred electron populations in IP \textit{p} states, as depicted in Fig. \ref{xanesxld}(h). Therefore, the XLD analysis points out that electron populated Ni and Mn 3\textit{d} orbitals will have significant overlaps with Ti 3\textit{d} orbitals across the interfaces through O 2\textit{p} orbitals and this may help in charge transfer from Ni/Mn to Ti cation sites, which is also corroborated with our isotropic XANES results.

These observations suggest that at the SB interface, Ti is in 3+ valence state. As we have observed previously \cite{SMajumder2019}, the unpaired 3\textit{d} electron from Ti$ ^{3+} $, can induce magnetism in BTO system by introducing magnetic exchange interactions through Ti$ ^{4-\epsilon} $-O-Ti$ ^{3+\epsilon} $-O-Ti$ ^{3+\epsilon} $-O-Ti$ ^{4-\epsilon} $ pathways. The coupling of spins at SNMO/BTO (or BTO/SNMO) interfaces would give rise to the observed exchange bias effect in SBS heterostructures. 

In summary, we have investigated the origin of unexpected magnetic behaviors by means of cooling field dependent EB and training characters in epitaxial SNMO/BTO/SNMO (SBS) heterostructure systems. Polarization dependent Raman spectroscopy and magnetization analysis confirm coexisting short scale antiferromagnetic disordered bonds and long range ferromagnetic ordered chains in SNMO layers. X-ray absorption measurements, configuration interaction simulations, and X-ray linear dichroism results reveal Ni/Mn 3\textit{d} to Ti 3\textit{d} orbital charge transfer across the SNMO/BTO interfaces, which can transform the BTO sandwich layer to possess magnetic moment at Ti sites at the interfaces. The interfacial coupling of moments at SNMO/BTO (or BTO/SNMO) interfaces eventually result in observed exchange bias effect in SBS heterostructures. \\ 

%\section*{Acknowledgments}
Authors acknowledge the Indus Synchrotron facility (India) for providing experimental access. S.M. thanks Dr. V. R. Reddy (CSR), and Mr. A. Gome (CSR) for RSM measurements; and Mr. Rakesh K. Sah (CSR) for the technical help in XAS measurements.

\bibliography{}
%\section*{REFERENCES}

\end{document}